\documentclass[%
 reprint,floatfix,
superscriptaddress,
bibnotes,
 amsmath,amssymb,
 aps,
]{revtex4-1}

\usepackage[usenames, dvipsnames]{color}

\usepackage[export]{adjustbox}
\usepackage{subfigure}
\usepackage{graphicx}
\usepackage{dcolumn}
\usepackage{bm}
\usepackage{hyperref}

\begin{document}

\preprint{APS/123-QED}

\title{Echoes from Quantum Black Holes}

\author{Qingwen Wang}
\email{qwang@pitp.ca}
\affiliation{Department of Physics and Astronomy, University of Waterloo, 200 University Ave W, N2L 3G1, Waterloo, Canada}
\affiliation{Waterloo Centre for Astrophysics, University of Waterloo, Waterloo, ON, N2L 3G1, Canada}
\affiliation{Perimeter Institute For Theoretical Physics, 31 Caroline St N, Waterloo, Canada}
 \author{Naritaka Oshita}%
 \email{noshita@pitp.ca}
    \affiliation{Perimeter Institute For Theoretical Physics, 31 Caroline St N, Waterloo, Canada}
 \affiliation{
  Research Center for the Early Universe (RESCEU), Graduate School
  of Science,\\ The University of Tokyo, Tokyo 113-0033, Japan}
  \affiliation{
   Department of Physics, Graduate School of Science, The University of Tokyo, Tokyo 113-0033, Japan}

 \author{Niayesh Afshordi}%
 \email{nafshordi@pitp.ca}
\affiliation{Department of Physics and Astronomy, University of Waterloo, 200 University Ave W, N2L 3G1, Waterloo, Canada}
\affiliation{Waterloo Centre for Astrophysics, University of Waterloo, Waterloo, ON, N2L 3G1, Canada}
\affiliation{Perimeter Institute For Theoretical Physics, 31 Caroline St N, Waterloo, Canada}




\date{\today}

\begin{abstract}
One of the most triumphant predictions of the theory if general relativity was the recent  LIGO-Virgo detection of gravitational wave (GW) signals produced in binary black hole (BH) mergers. However, it is suggested that
exotic compact objects, proposed in quantum gravity models of BHs, may produce similar classical GW waveforms, followed by delayed repeating ``echoes''. In a companion paper \cite{Oshita:2019sat}, we have presented different arguments for a universal Boltzmann reflectivity of quantum BH horizons. Here, we investigate the resulting echoes from this prescription. We derive corresponding quasi-normal modes (QNMs) for quantum BHs analytically, and show how their initial conditions can be related to the QNMs of classical BHs. Ergoregion instability is suppressed by the imperfect reflectivity. We then compare the analytic and numerical predictions for echoes in real time, verifying their consistency. In particular, we find that the amplitudes of the first $\sim 20$ echoes decay inversely with time, while the subsequent echoes decay exponentially.  Finally, we present predictions for the signal-to-noise ratio of echoes for spinning BHs, which should be imminently detectable for massive remnants, subject to the uncertainty in the nonlinear initial conditions of the BH merger. 
\end{abstract}

\pacs{Valid PACS appear here}
\maketitle


\section{\label{sec1}Introduction}

Attempts to solve black hole (BH) information paradox \cite{Lunin:2001jy, Lunin:2002qf, Mathur:2005zp, Mathur:2008nj, Mathur:2012jk,Braunstein:2009my, Almheiri:2012rt} and cosmological constant problems \cite{PrescodWeinstein:2009mp} suggest that non-perturbative quantum gravitational effects may lead to Planck-scale modifications of the classical BH horizons. Markedly, proposals like gravastars \cite{Mazur:2004fk}, fuzzballs \cite{Lunin:2001jy, Lunin:2002qf, Mathur:2005zp, Mathur:2008nj, Mathur:2012jk}, aether BHs \cite{PrescodWeinstein:2009mp}, and firewalls \cite{Braunstein:2009my, Almheiri:2012rt} amongst others \cite{Barcelo:2015noa, Kawai:2017txu,Giddings:2016tla} all modify the standard structure of the BH stretched horizons with a non-classical surface (or structure).  

While these attempts have been largely limited to theoretical speculation, the LIGO-Virgo collaboration reported unprecedented gravitational wave (GW) observations from binary BH merger events in 2016 \cite{TheLIGOScientific:2016agk, TheLIGOScientific:2016pea, Abbott:2016blz, Abbott:2016nmj, Abbott:2017vtc, Abbott:2017oio, TheLIGOScientific:2017qsa, Abbott:2017gyy}, providing a possible route to test these ideas. In particular,  \cite{Cardoso:2016rao, Cardoso:2016oxy,Cardoso:2017njb, Cardoso:2017cqb} argued that GW observations can shed light on the true structure of BH horizons, as any modification could produce delayed GW ``echoes'' of the merger event. Followed by this idea, two independent groups reported tentative evidence, at 2-3$\sigma$ level, for presence of echoes in LIGO observations of different binary BH mergers \cite{abedi2016echoes,Conklin:2017lwb}, although this interpretation remains controversial \cite{Westerweck:2017hus,Abedi:2018pst}. More recently, \cite{BNS} found a tentative detection of (lower harmonics of) echoes, at $4.2\sigma$ level, from a highly spinning ``black hole'' remnant in the aftermath of the GW170817 binary neutron star merger. Besides, a mass shell outside the classical BHs also generate the similar echoes, but \cite{Konoplya:2018yrp} discuss that the echoes from modifications of the horizon dominate the signal. 

Existing studies consider both phenomenological echo templates (e.g., \cite{abedi2016echoes,Maselli:2017tfq}), and those that come from solving (linearized) Einstein equations with modified boundary conditions. The latter, so far, have mostly focused on Schwarzschild BHs (e.g.,  \cite{Cardoso:2016rao,Cardoso:2016oxy, Price:2017cjr, Mark:2017dnq, Volkel:2018hwb, Volkel:2017kfj}), although more attempts now extend this to Kerr metric as realistic BHs have spin \cite{Nakano:2017fvh, Wang:2018gin, Maggio:2018ivz,Burgess:2018pmm}. One major drawback is that these studies often assume a perfectly reflective, or otherwise ad hoc boundary condition (but see \cite{Cardoso:2019apo} for a proposal based on BH area quantization). A more physical boundary condition is urgently needed for a more realistic echo template, as abundant GW data is on its way. Moreover, the choice of boundary condition can determine the (in)stability of the ergoregion in compact objects \cite{1978CMaPh..63..243F, Cardoso:2007az,Cunha:2017qtt,Maggio:2017ivp}.

In a companion paper, we study this question assuming that BHs are quantum systems that follow standard rules of quantum mechanics and thermodynamics \cite{Oshita:2019sat}. There, we found that independent arguments based on thermodynamic detailed balance, fluctuation-dissipation theorem, and CP-symmetry of the extended BH spacetime, remarkably all lead to a universal  Boltzmann energy flux reflectivity: 
\begin{equation}
    \frac{E_{\rm out}}{E_{ \rm in}} = \exp\left(-\frac{\hbar |\tilde{\omega}|}{k_B T_{\rm H}}\right), \label{Boltzmann}
\end{equation}
where $\tilde{\omega}$ is the horizon-frame frequency and $T_{\rm H}$ is the Hawking temperature.  In this paper, we investigate the quasinormal modes (QNMs) of these quantum BHs, and show how their excitation can be related to QNMs of classical BHs. This result can be used to make predictions for GW echoes from quantum BHs, which we verify using numerical and analytic calculations. We further study the detectability of these echoes, and show that ergoregion instability is suppressed, consistent with astrophysical \cite{Narayan:2013gca} and GW observations \cite{Barausse:2018vdb}. 



We organize this paper as follows: Sec. \ref{sec2} calculates the QNMs from a Boltzmann boundary condition analytically, using the tools developed in \cite{Maggio:2018ivz}. Next, Sec. \ref{sec3} presents echoes in the time domain both numerically and analytically, and confirms the QNMs calculated in Sec.\ref{sec2} with numerical results. Also the importance of the initial condition is manifested in the time domain. Then, in Sec. \ref{sec4}, we discuss how the ergoregion instability is quenched, and draw conclusions in Sec. \ref{sec5}.

If not specified, we use $G = k_{\rm B} = c= \hbar =1$. $\tilde{\omega}$ is the near horizon frequency while $\omega$ is the frequency at infinity. For concreteness, we use the best fit properties and waveforms resulting from the GW150914 merger event, provided by the LIGO-Virgo collaboration \cite{TheLIGOScientific:2016agk, TheLIGOScientific:2016pea} \footnote{https://losc.ligo.org/events/GW150914/}. In particular, unless mentioned otherwise, the detector frame mass and reduced spin parameter of the remnant used for the echo calculation are $M_{\rm fin} = 67.6~ M_{\odot}$ and  $\bar{a} \equiv a/M_{\rm fin} = 0.67$, respectively. 

\section{Reflectivity of Quantum Black Holes}\label{FD_summary}

Let us start by reviewing the results of our companion paper: In \cite{Oshita:2019sat}, we introduced the notion of reflectivity of quantum black hole horizons, and provided physical arguments for why it should be non-vanishing for classical (i.e. large amplitude) GWs. In particular, we provided three independent derivations based on detailed balance, fluctuation-dissipation theorem, and CP-symmetry of BH extended spacetime to derive the Boltzmann energy flux reflectivity (\ref{Boltzmann}). Out of these three, the fluctuation-dissipation theorem gives the most explicit result for both the amplitude and phase of reflected GWs, and thus we shall summarize it here.

Based on the fluctuation-dissipation theorem, the interaction of any single degree of freedom with a thermal bath can be approximated as a combination of a dissipation/friction and a stochastic force \cite{1966RPPh...29..255K}. The two effects only balance each other when the (statistical distribution of the) degree of freedom reaches the same temperature as the thermal bath. For our derivation, we then looked at the amplitude of GW modes near BH horizons. For large amplitudes of GWs, we expect the fluctuation term to be negligible, but the dissipation/friction term, which scales with the ``velocity'' would stay relevant.  Furthermore, we assume that the relative effect of dissipation to scale as the gravitational interaction strength $\Omega/E_{\rm Pl}$, where $\Omega$ is the proper frequency of the mode and $E_{\rm Pl}$ is the Planck energy. This yields a modified wave equation near BH horizon
\begin{equation}
    \left[ -i \frac{\gamma \Omega (x)}{E_{\text{Pl}}} \frac{d^2}{dx {}^2} + \frac{d^2}{dx {}^2} + \tilde{\omega}^2 \right]
\psi_{\omega} (x) \simeq 0, \label{FD_eq}
\end{equation}
where $\tilde{\omega}$ is the horizon-frame frequency, and $\Omega(x) = \sqrt{g^{00}(x)} \tilde{\omega} \simeq e^{-\kappa x} \tilde{\omega}$. Furthermore, $x$ is the tortoise coordinate that approaches $-\infty$ at horizon, $\kappa = 2\pi T_{\rm H}$ is the surface gravity, and $\gamma$ is a dimensionless parameter controlling the strength of dissipative coupling. 

Equation (\ref{FD_eq}), with reasonable boundary conditions at infinities, has a unique analytic solution in terms of a hypergeometric function \cite{Oshita:2019sat}. Any incoming wave at $x \rightarrow \infty$ is then partially reflected and absorbed where $\gamma \Omega(x) \sim E_{\rm Pl}$. Using the asymptotic behavior of the hypergeometric function, we find that the ratio of outgoing to incoming wave amplitude at $x=0$ is given by:
\begin{align}
R_{\rm wall} &\equiv \frac{A_{\rm out}}{A_{\rm in}} = \frac{\left( \gamma \tilde{\omega} \right)^{-2i \tilde{\omega}/ \kappa} \Gamma (-2i \tilde{\omega}/ \kappa)\Gamma (i \tilde{\omega}/\kappa) \Gamma (1+i \tilde{\omega}/\kappa)}{\Gamma (-i \tilde{\omega}/\kappa) \Gamma (1-i \tilde{\omega}/\kappa)\Gamma (2i \tilde{\omega}/ \kappa)} \nonumber \\
&\simeq e^{- \frac{\tilde{\omega}}{ 2 T_{\rm H}}} { \left( \gamma \tilde{\omega}\right)}^{ -\frac{ i \tilde{\omega}}{  \pi T_{\rm H}}}, ~{\rm for}~~ \tilde{\omega} \ll \kappa=2\pi T_{\rm H}, \label{BC_eq}
\end{align}
where we have now used Planck units $E_{\rm Pl} = G^{-1/2} =1$. Note that the square of the absolute value of this ratio gives the Boltzmann flux reflectivity (\ref{Boltzmann}), while the phase indicates the approximate location of reflection at $x \simeq \kappa^{-1} \ln|\gamma\tilde{\omega}|$. We will then use this boundary condition in the next section to find the QNMs of quantum BHs. 

\section{\label{sec2} Quasinormal modes}

In this section, we investigate the QNMs based on the fluctuation-dissipation theorem for quantum BHs, that we introduced in \cite{Oshita:2019sat}, and have summarized in Section \ref{FD_summary} above. We use two analytic methods: the geometric optics approximation, or the asymptotic matching method based on \cite{Maggio:2018ivz}, which both yield the same analytic formula. 

We introduce the Newman-Penrose (NP) Formalism which greatly simplifies the perturbation in the Kerr metric, reducing it to only a single master equation known as the Teukolsky equation (see \citeauthor{teukolsky1973perturbations} \cite{teukolsky1973perturbations} for details):
\begin{widetext}
\begin{eqnarray}\label{eq:teuk} 
\left[\frac{(r^2+a^2)^2}{\Delta}-a^2 \sin ^2 \theta\right] \frac{\partial^2 \psi}{\partial t^2}+ \frac{4Mar}{\Delta}\frac{\partial^2 \psi}{\partial t \partial \varphi}+\left(\frac{a^2}{\Delta}-\frac{1}{\sin^2 \theta}\right)\frac{\partial^2 \psi}{\partial \varphi^2}-\Delta^{-s} \frac{\partial}{\partial r} \left(\Delta^{s+1} \frac{\partial \psi}{\partial r}\right)-\frac{1}{\sin \theta} \frac{\partial}{\partial \theta}\left(\sin \theta \frac{\partial \psi}{\partial \theta}\right)\nonumber\\-2s\left[\frac{a(r-M)}{\Delta}+\frac{i \cos \theta}{\sin^2\theta}\right]\frac{\partial\psi}{\partial \varphi}-2s\left[\frac{M(r^2-a^2)}{\Delta}-r-ia\cos\theta\right]\frac{\partial \psi}{\partial t}+(s^2\cos^2\theta-s)\psi= 0, ~ 
\end{eqnarray}
\end{widetext}
where the fields $\psi$ for each spin weight $s$ corresponds to NP quantities presented in Table \ref{master}. 

\begin{table}
\caption{\label{master}%
Corresponding fields $\psi$ for different spin weight $s$ in the Master equation. Here $\rho^{-1}=-(r-i a \cos\theta)$
}
\begin{ruledtabular}
\begin{tabular}{c|cccc}
\textrm{s}&
\textrm{0}&
\textrm{-1/2, 1/2}&
\textrm{-1, 1}&-2, 2\\
\colrule
$\psi$ & $\Phi$ & $\chi_0, \rho^{-1} \chi_1$ & $\phi_0, \rho^{-2} \phi_2$ & $\Psi_0, \rho^{-4} \Psi_4$\\
\end{tabular}
\end{ruledtabular}
\end{table}

In the frequency domain, the Teukolsky equation (\ref{eq:teuk}) is separable in coordinates and can be decomposed into 4 ODEs. Furthermore, the symmetries in time and azimuth, allow for Fourier space decomposition in $t$ and $\varphi$:   
\begin{widetext}
\begin{eqnarray}
&&\psi=\frac{1}{2\pi} \int d\omega e^{i(-\omega t +m \varphi)} S[\theta] R[r],\label{R}\\
&&\Delta^{-s} \frac{d}{dr}	\left( \Delta^{s+1} \frac{dR}{dr} \right)+\left[ \frac{K^2-2is(r-M)K}{\Delta}+4is\omega r -\lambda \right]R=0,\label{r}\\
&&\frac{1}{\sin\theta}\frac{d}{d\theta}\left(\sin\frac{dS}{d\theta}\right)+\left(a^2 \omega^2 \cos^2\theta-\frac{m^2}{\sin^2\theta}-2a\omega s \cos\theta-\frac{2ms\cos\theta}{\sin^2\theta}-s^2 \cot^2\theta+s+A_{slm}\right)S=0,\label{s}
\end{eqnarray}
\end{widetext}
where $K=(r^2+a^2)\omega -am$ and $\lambda=A_{slm}+a^2 \omega ^2 -2am\omega$. The solution for the angular mode is the spin-weighted spheroidal harmonic (full discussion can be found in \cite{Berti:2005gp}). For the radial equation, we introduce Detweiler’s function \cite{Detweiler:1977gy}
\begin{eqnarray}
{}_sX_{lm} = \Delta (r^2 +a^2)^{1/2} \left[ \alpha {}_sR_{lm} + \beta \Delta ^{s+1} \frac{d{}_sR_{lm}}{dr}\right],
\end{eqnarray}
where $\alpha$ and $\beta$ are radial functions and the different choices of them influence the $V(r,\omega)$ in Eq. (\ref{SN}). The radial master equation becomes a simple non-singular wave equation with two independent asymptotic solutions $X^+_s$ and $X^-_s$, where we omit indices $l$ and $m$:
\begin{eqnarray}
\label{SN}
&\frac{d^2 {}_sX_{lm}}{ dx^2} - V(r,\omega) {}_sX_{lm}=0,\\
&X^+_s=\left\{
\begin{array}{rcl}
B_+ e^{-i \tilde{\omega} x} ,      &      & x     \rightarrow     - \infty\\
e^{+i \omega x}+ A_+ e^{-i \omega x} ,   &      & x     \rightarrow     \infty\\
\end{array} \right. \\
&X^-_s=\left\{
\begin{array}{rcl}
e^{+i \tilde{\omega} x}+ A_- e^{-i \tilde{\omega} x} ,      &      & x     \rightarrow     - \infty\\
B_- e^{+i \omega x},   &      & x     \rightarrow     \infty\\
\end{array} \right. 
\end{eqnarray}
where $x$ is the tortoise coordinate (defined as $x=\int \frac{r^2+a^2}{r^2-2Mr +a^2} dr$, approaching -$\infty$ at horizon), while $\tilde{\omega}= \omega-\frac{a m}{2Mr_{+}}$ and $r_+ = M+\sqrt{M^2-a^2}$. The potential $V(r,\omega)$ can be found in \cite{Maggio:2018ivz}.

Now, we apply the fluctuation-dissipation theorem \cite{Oshita:2019sat}. With the modified Einstein equation from the theorem (\ref{FD_eq}), the boundary condition obtained for the asymptotic solution $X^-_s$ near horizon is fixed by Eq. (\ref{BC_eq}):
\begin{equation}
A_{-}=  R^{-1}_{\rm wall} = e^{+ \frac{\tilde{\omega}}{ 2 T_{\rm H}}} { \left( \gamma \tilde{\omega}\right)}^{ \frac{ i \tilde{\omega}}{  \pi T_{\rm H}}},
\end{equation}
where $\gamma$ was the free parameter that quantified dissipation in the fluctuation-dissipation theorem. We further assume that the imaginary part of frequency is much smaller than its real part, thus
\begin{eqnarray}
&A_{-} \simeq  e^{+ \frac{|\tilde{\omega}|}{ 2 T_{\rm H}}+ \frac{ i \tilde{\omega}}{  \pi T_{\rm H}} \ln(\gamma | \tilde{\omega} | ) }. 
\end{eqnarray}

We can compare this result to \cite{Maggio:2018ivz} with a Neumann boundary condition,  $\frac{ d X_s^- }{d x}=0 $, at $r_0= r_+ (1+\epsilon)$: They find $ A_{-} = e^{2 i \tilde{\omega} x_0} $, with $x_0=x(r_0)$. We can thus identify ${ \left( \gamma \tilde{\omega}\right)}^{ \frac{ i \tilde{\omega}}{  \pi T_{\rm H}}} = e^{2 i \tilde{\omega} x_0}$ or 
\begin{equation}
 x_0 \equiv \frac{ \ln(\gamma | \tilde{\omega}|) }{2 \pi T_{\rm H}},   
\end{equation}
as the effective position of the reflecting wall. Moreover, as discussed above, the energy flux reflectivity of the wall is exactly given by a Boltzmann factor $ e^{- \frac{|\tilde{\omega}|}{  T_{\rm H}}} $. 

Since the effective position of the wall changes very slowly as the $\ln(\tilde{\omega})$ (for $\tilde{\omega} \ll 1$ in Planck units), it can be translated to an approximately constant time delay between subsequent echoes:
\begin{equation}
    \Delta t_{\rm echo} \equiv 2|x_0| = -\frac{ \ln(\gamma | \tilde{\omega}|) }{\pi T_{\rm H}}.
\end{equation}

\begin{figure}
\hspace*{-2.5cm} 
\includegraphics[width=0.7\textwidth]{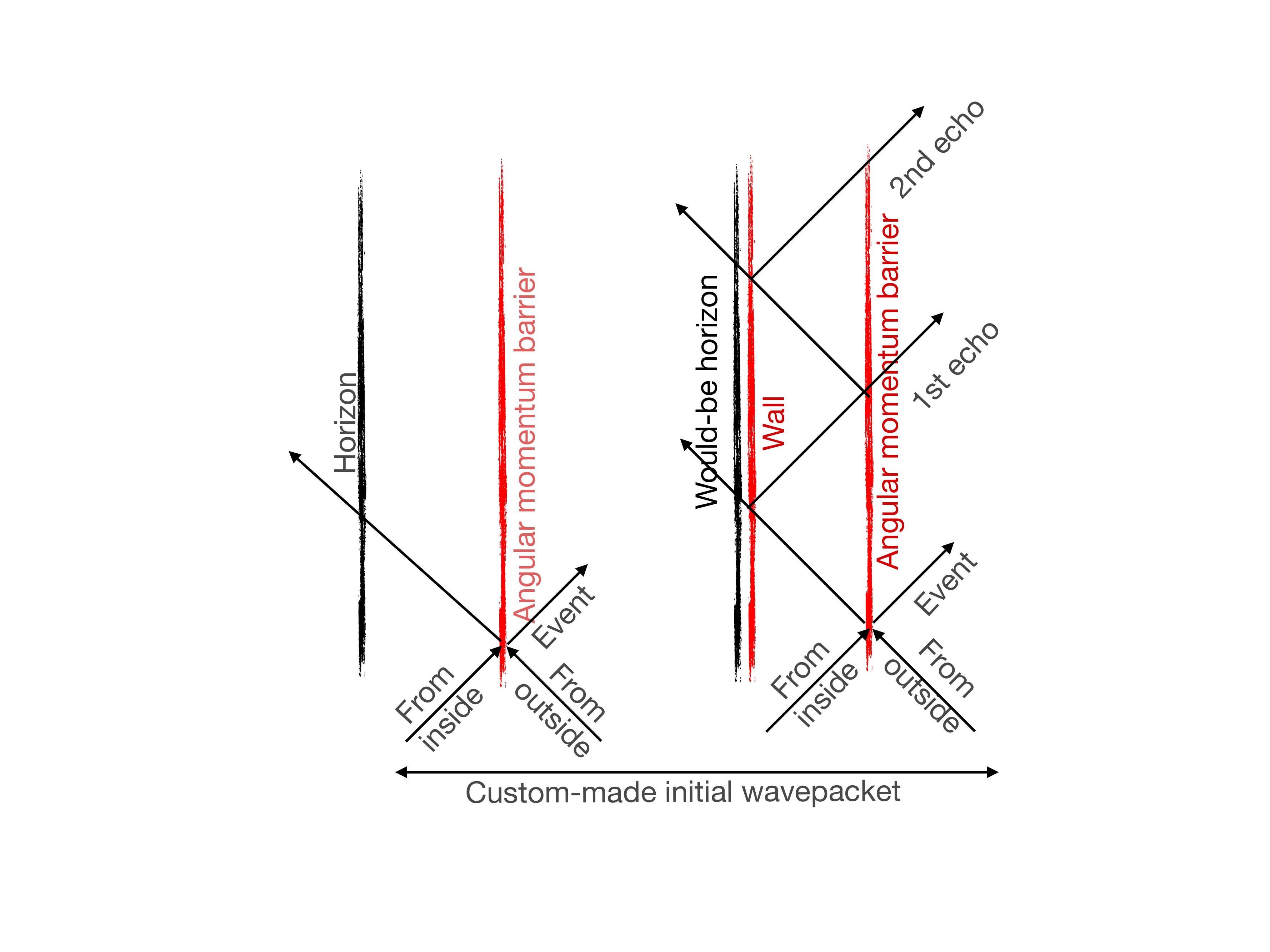}
\caption{\label{geo} Propagation of ingoing/outgoing wavepacket in a classical (left) and quantum (right) BH spacetime . For classical BH, angular momentum barrier reflects (transmits) low (high) frequency modes while the higher (lower) frequencies cross (reflect off) the barrier and fall through the horizon. For quantum BHs,  low frequency modes are trapped between the quantum wall and the angular momentum barrier, slowly leaking out as repeating echoes.}
\end{figure}

Let us now find the QNMs for the quantum BH. Fig. \ref{geo} shows the geometric ``optics'' picture for the echology, which is valid as long as $|\tilde{\omega}| \gg |x_0|^{-1}$. In this limit, we can obtain the quantum BH response by using $R_{\rm BH} $ and $T_{\rm BH} $($R^*_{\rm BH} $ and $T^*_{\rm BH} $), the reflectivity and transmissvity of classical BHs' angular momentum barrier with an ingoing (outgoing) wavepacket from outside (inside):
\begin{eqnarray}
\left({\frac{h_{\rm out}}{h_{\rm in}}}\right)_{\rm outside} = &  R_{\rm BH}+ \sum_{n=1}^\infty |T_{\rm BH}|^2 R_{\rm wall}^n  R^{*(n-1)}_{\rm BH} e^{- 2i n \tilde{\omega} x_0} \nonumber\\
 = & R_{\rm BH} + \frac{| T_{\rm BH}|^2 R_{\rm wall}  e^{- 2 i \tilde{\omega} x_0}}{1-R_{\rm wall} R^*_{\rm BH}e^{- 2 i \tilde{\omega} x_0} } \label{outside} \\
\left({\frac{h_{\rm out}}{h_{\rm in}}}\right)_{\rm inside} = &  T^*_{\rm BH}+ \sum_{n=1}^\infty T_{\rm BH}^* R_{\rm wall}^n  R^{*n}_{\rm BH}  e^{- 2i n \tilde{\omega} x_0}   \nonumber\\
 = & T_{\rm BH}^* + \frac{ T_{\rm BH}^*  R^{*}_{\rm BH} R_{\rm wall}  e^{- 2 i \tilde{\omega} x_0}}{1-R_{\rm wall} R^*_{\rm BH}e^{- 2 i \tilde{\omega} x_0} }
 \label{inside}
\end{eqnarray}

The first term of each equation is the initial observed event as in Fig. \ref{geo}, which is the same for classical BHs and quantum BHs. The subsequent terms in Eqs. (\ref{outside})  and  (\ref{inside}) represent the first echo, second echo, etc., which can be summed as a geometric series. The QNMs are poles of the response function, or the zero's of the denominator, $1-R_{\rm wall} R^*_{\rm BH}e^{- 2 i \tilde{\omega} x_0}=0$, where $R_{ \rm wall} = e^{-\frac{ |\tilde{\omega}|}{2 T_{\rm H }}}$ for our quantum BHs. Near $ \tilde{\omega} \simeq 0$, we have the least-damped modes, which we shall focus on next. We numerically confirm $R^*_{\rm BH} \simeq \pm 1$ for $ \tilde{\omega} \ll T_{\rm H}$, where plus (minus) is for $s=-1~ (s=0,-2)$. Hence, $\tilde{\omega}_q$ for QNMs satisfy:
\begin{eqnarray}
\label{bc}
& e^{- 2 i \tilde{\omega} x_0  -\frac{ |\tilde{\omega}|}{T_{\rm H }} } = \pm 1,\\
& \tilde{\omega}_n= \frac{ q \pi}{2 x_0} \left[1 - \frac{{\rm sgn}(q) \times i}{4 x_0 T_{\rm H}}\right] \label{qnm_freq},
\end{eqnarray}
where $q=2n+1$ for $s=0, -2$, and $q=2n$ for $s=-1$, with $n \in \mathbb{Z}$. We arrive at the same result via the asymptotic matching method used in \cite{Maggio:2018ivz}. Since we prove that the ratio of outgoing and ingoing waves of solution of Eq. (\ref{r}) (denoted as $\frac{C_1}{C_2}$ in \cite{Maggio:2018ivz}) is proportional to $ A_{-}^{-1} $, just simply multiplying the extra Boltzmann reflectivity by Eqs. (A9) and (A13) in \cite{Maggio:2018ivz} recovers Eq. (\ref{bc}), hence the QNMs.

\section{Real time Echoes}

QNMs are crucial to the structure of echoes in the real time. Our analytic derivation of QNMs, in Sec. \ref{sec2} above,  is only valid for $|\tilde{\omega}| \ll T_{\rm H}$, but might be sufficient to encode information for the real time echoes since the least-damped mode is in the same range. In this section, we calculate the echoes numerically in the geometric optics limit (Eqs. (\ref{outside}) and (\ref{inside})), and analytically from the QNMs found in Sec \ref{sec2}, confirming that two calculations are consistent.
\label{sec3}

\subsection{Numerical Echoes from geometric optics approximation}

While the realistic behavior of echoes should come from the nonlinear evolution, starting with two inspiraling BHs, we can imitate this by linear initial conditions with a wavepacket hitting the angular momentum barrier, from inside {\it or} outside, producing exactly the same ringdown waveform as in the LIGO template for GW150914, denoted as $h_{\rm LIGO}$. We can then use Eqs. (\ref{outside}) and (\ref{inside}) to predict echo waveform, using linear initial conditions and geometric optics limit, in frequency space:
\begin{eqnarray}
&&h_{\rm out} = h_{\rm LIGO} \left(1 + \frac{  {\cal M}_{\rm initial} R_{\rm wall}  e^{- 2 i \tilde{\omega} x_0}}{1-R_{\rm wall} R^*_{\rm BH}e^{- 2 i \tilde{\omega} x_0} } \right), \label{num_echo}\\
&&R_{\rm wall} = \exp\left(-\frac{\tilde{\omega}}{2T_{\rm H}}\right), {\cal M}_{\rm initial} = \left\{
\begin{array}{rcl}
\frac{|T_{\rm BH}|^2}{R_{\rm BH}},      &      &\textrm{from outside,}\\
R_{\rm BH}^*,   &      &                        \textrm{from inside.}\\
\end{array} \right. \nonumber\\
\label{M}
\end{eqnarray}
 
\begin{figure}
\hspace*{-0.5cm} 
\includegraphics[width=0.5\textwidth]{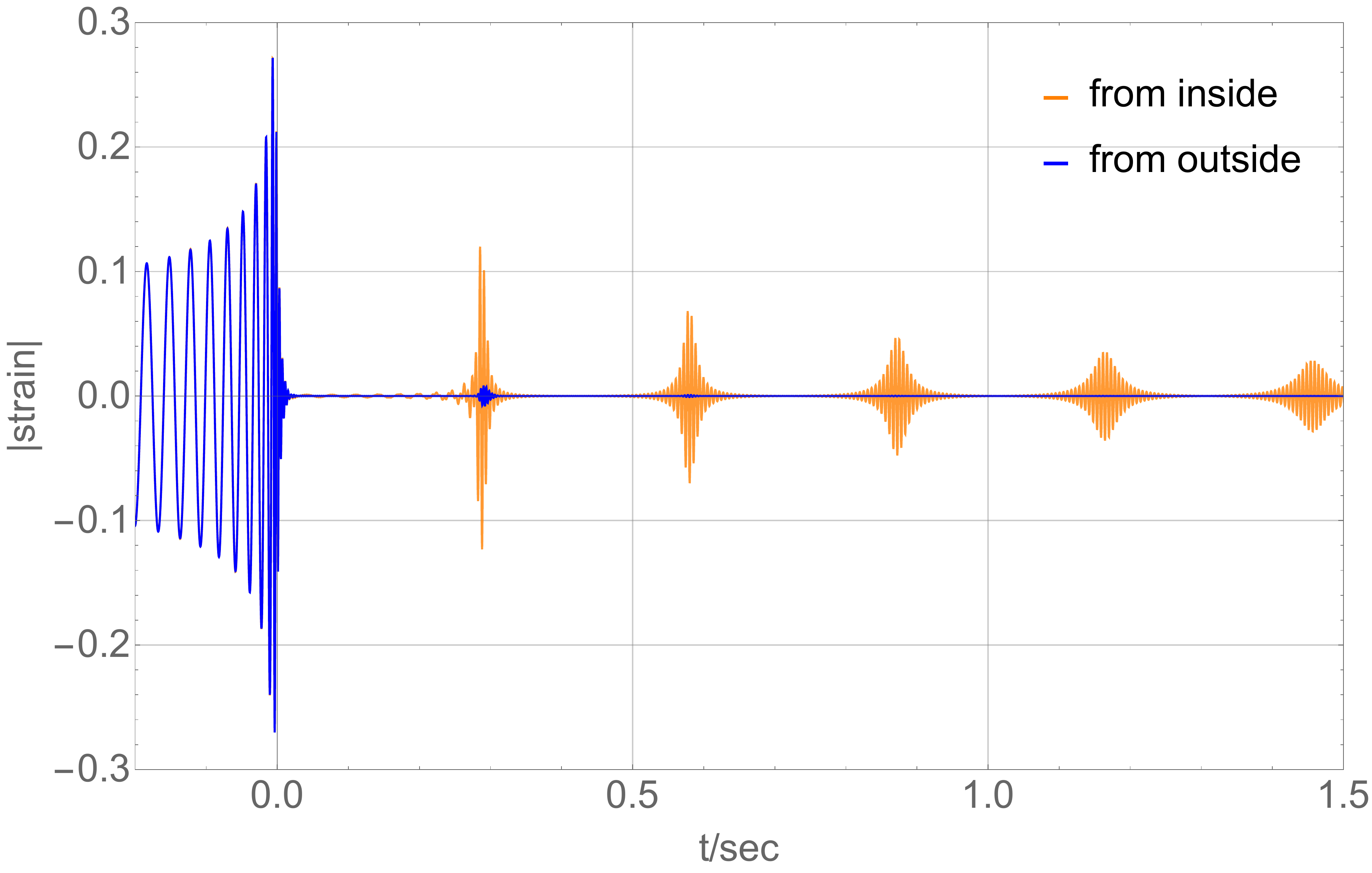}
\caption{\label{echoes} Real time echoes from the geometric optics approximation, and $\gamma \sim 1$. The first burst is exactly the same as in the LIGO template for GW150914 (only with strain rescaled for comparison with the analytic solution in Fig. \ref{compare}). Purple (orange) is for the initial wavepacket coming from outside (inside).}
\end{figure}
 
 Reflectivity and transmissivity of classical BHs can be found numerically by solving the Teukolsky equation \footnote{Here, we use the numerical solutions  from \citeauthor{Conklin:2017lwb} \cite{Conklin:2017lwb}.}. Fig. \ref{echoes} shows the prediction for real-time echo waveforms, by Fourier transforming Eq. (\ref{num_echo}). By construction, the first burst has the exact same waveform as the LIGO template for GW150914 (note that we rescale the strain for later comparison with analytic result in Fig. \ref{compare}). The outside initial condition produces smaller echoes than the inside since reflection rate of BH is near 1 around the main frequency (around $\tilde{\omega} \sim 0$, where $R_{\rm wall} \sim 1$), and the transmission is around 0. Hence, ${\cal M}_{\rm initial}$ in Eq. (\ref{M}) is much smaller for the outside condition, and so are the echoes. Another feature is that the echoes become broader over time, since higher frequencies leak more rapidly.

\subsection{Analytic Echoes from QNMs}

QNMs are the pure outgoing solution as $X^-_s$ in Sec \ref{sec2}. Hence, we should also be able to recover the numerical real-time echo solution with the analytic QNMs calculated in Sec \ref{sec2}. We assume that the solution is a sum over QNMs: 
\begin{equation}
h_{\rm out}(t) \simeq \sum_{n=-\infty}^\infty B_n e^{- i (\tilde{\omega}_n+\frac{a}{M r_+}) t},\label{qnm_expansion}
\end{equation}
where $B_n$'s are the complex amplitudes of the QNMs, and we use $l=m=2$ for the dominant QNMs.

For event GW150914, the classical ringdown is well-modelled by a single dominant QNM (or a Lorentzian template) with $\omega_* \simeq 1470 - i 250$ {\rm rad/s} \cite{TheLIGOScientific:2016src}:
\begin{eqnarray}
&&h_{\rm Lorentz}(t) = \Theta(t) e^{- i \omega_* t} = \frac{1}{2i \pi}\int \frac{e^{-i\omega t}}{\omega-\omega_*} d\omega  \nonumber\\
\label{int}
&\simeq& \frac{1}{2 i x_0}\int \frac{1}{\tilde{\omega}_n+\frac{a}{M r_+} - \omega_*} e^{- i (\tilde{\omega}_n+\frac{a}{M r_+}) t} dn,\label{lorentz}
\end{eqnarray}
where we used Eq. (\ref{qnm_freq}) to approximate $\tilde{\omega}_n$, and ignored the imaginary part of $\tilde{\omega}_n$ \footnote{Note that $\left|\frac{\Im \tilde{\omega}}{\Re \tilde{\omega}}\right| = -\frac{\pi}{2 \ln(\gamma |\tilde{\omega}|)} \ll 1$  }.

Now, comparing Eq. (\ref{qnm_expansion}) and Eq. (\ref{lorentz}), we notice that the dominant QNM of the classical BH can be simply written as the sum over the QNMs of the quantum BH, by replacing $\int dn \to \sum_n$: 
\begin{equation}
h_{\rm out}(t) \simeq \sum_{n=-\infty}^\infty \frac{e^{- i (\tilde{\omega}_n+\frac{a}{M r_+}) t}}{2ix_0 \left(\tilde{\omega}_n+\frac{a}{M r_+} - \omega_*\right) },\label{qnm_lorentz}
\end{equation}
In other words, we assume that, in the $x_0 \to -\infty$ limit, the classical and quantum BHs have identical waveforms. However, for finite $x_0$, if we ignore the imaginary part of $\tilde{\omega}_n$'s, all QNMs have a common period of $\Delta t_{\rm echo} =2 |x_0|$, leading to periodic echoes after this time. However, the fact that $\Im \tilde{\omega} <0$ implies that subsequent echoes will decay.

\begin{figure}
\hspace*{-0.5cm}
\includegraphics[width=0.49\textwidth]{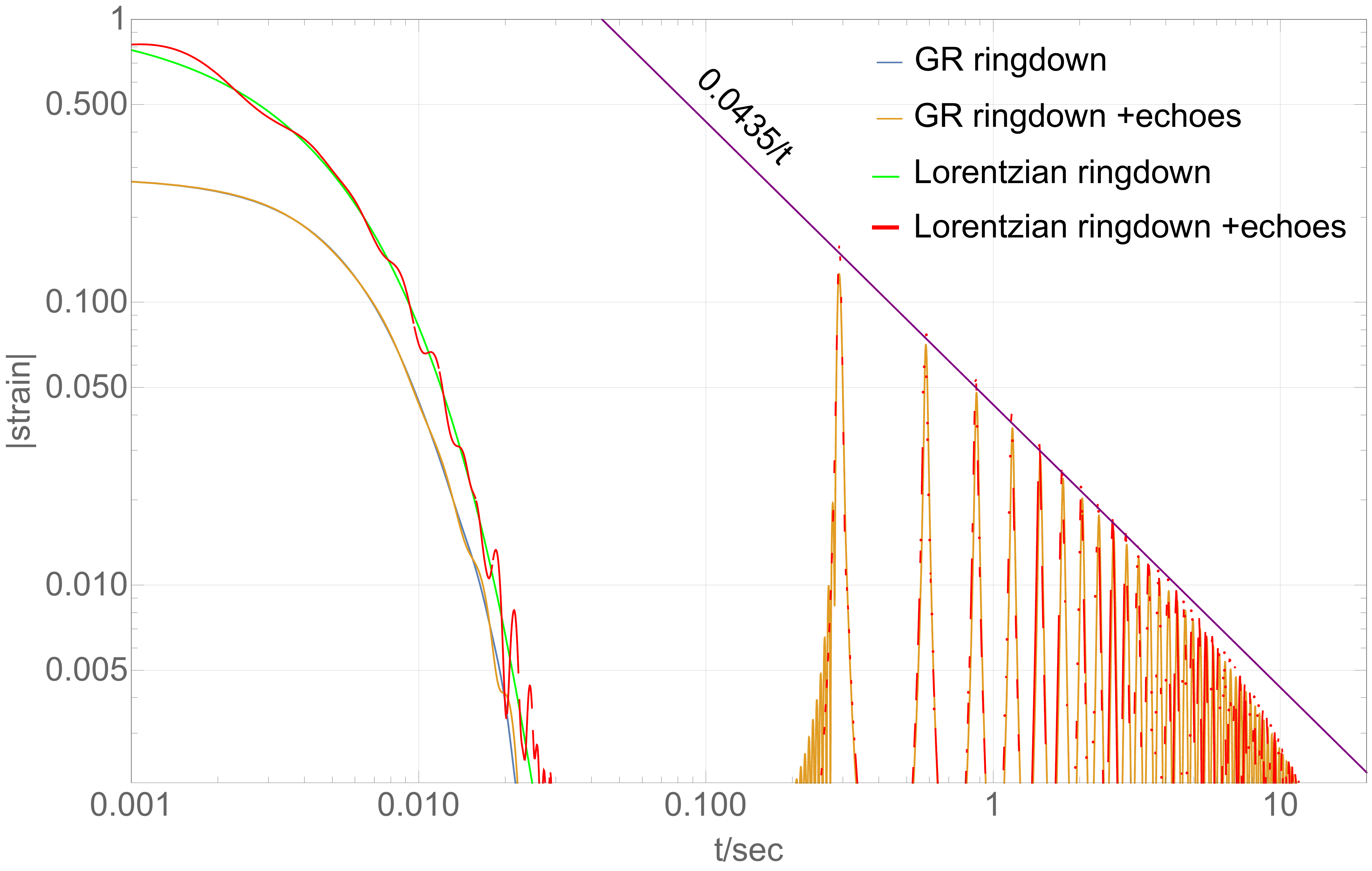}
\caption{\label{compare} The real-time echoes from the geometric optics approximation applied to GR template for GW150914 (same as orange curve in Fig. \ref{geo}), compared to the Lorentzian analytic approximation of QNMs. Note that the amplitude of the GR template is rescaled to make the first echoes match. The echoes initially decay as $1/t$, as many QNMs contribute to echoes. However, after $\sim 20$ echoes, only the least-damped QNM survives and thus strain starts to decay exponentially.}
\end{figure}

Fig. \ref{compare} compares the analytic prediction from Eq. (\ref{qnm_lorentz}) with the numerical result from (Fourier transform of) Eq. (\ref{num_echo}), for a wavepacket coming from inside the barrier (which is expected to be expandable in terms of quantum BH QNMs). The red dashed curve is the analytic solution, which matches very well with the orange curve from the numerical calculation. They both decay as $1/t$ at the beginning (first $\sim 20$ echoes), but then start to fall off exponentially. Note that we rescale the amplitude of LIGO template for GW150914 in the numerical calculation, to match the first echoes in both numerical and analytic solutions \footnote{Hence, if we rescale them to have the same initial event, then the numerical echoes are larger than analytic. This is due to the fact that in the analytic Lorentzian template, we only use a single QNM of the classical BH, but we use the full LIGO template in the numerical solution.}.
\begin{figure*}
\includegraphics[width=0.3 \textwidth]{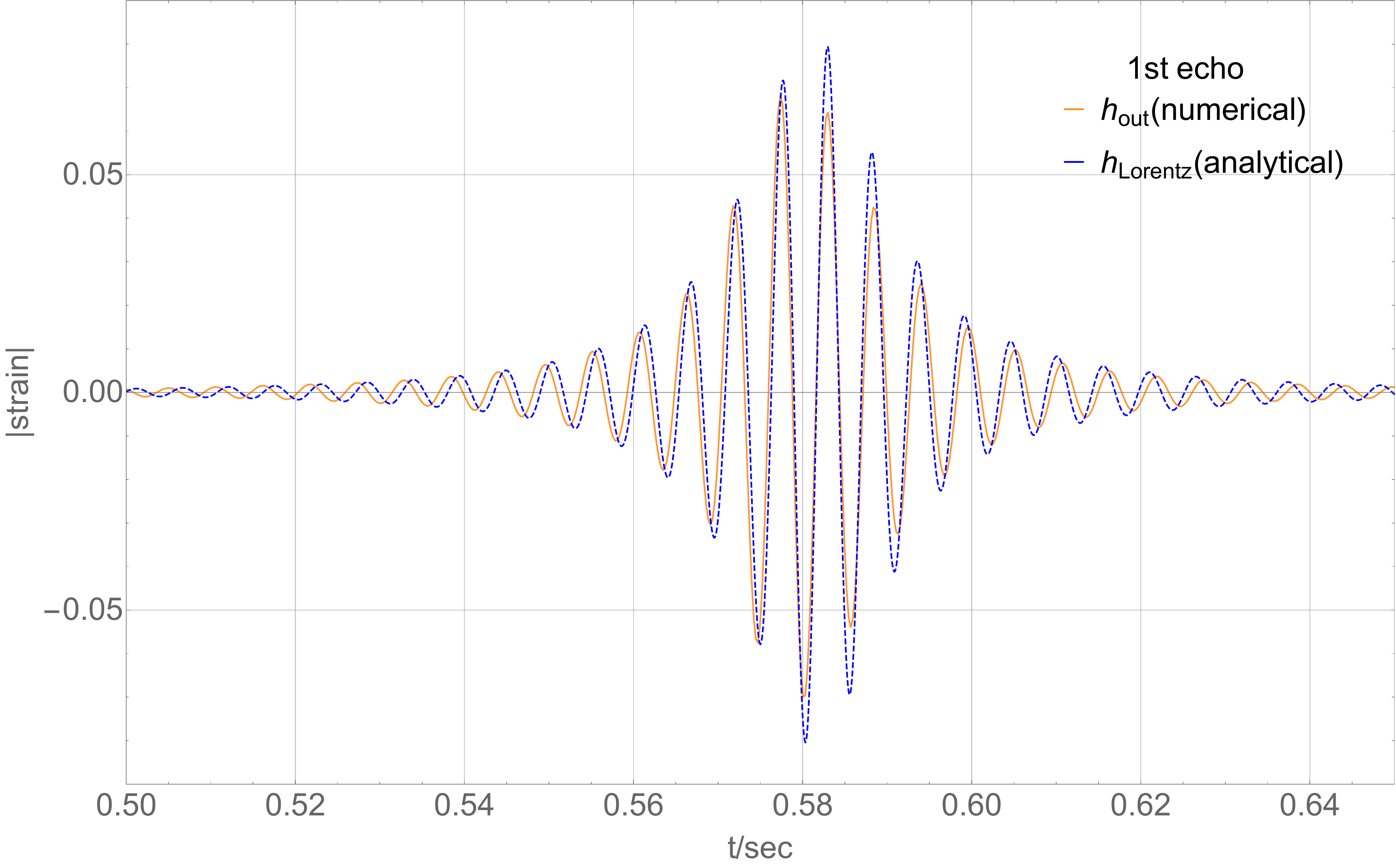}
\includegraphics[width=0.3 \textwidth]{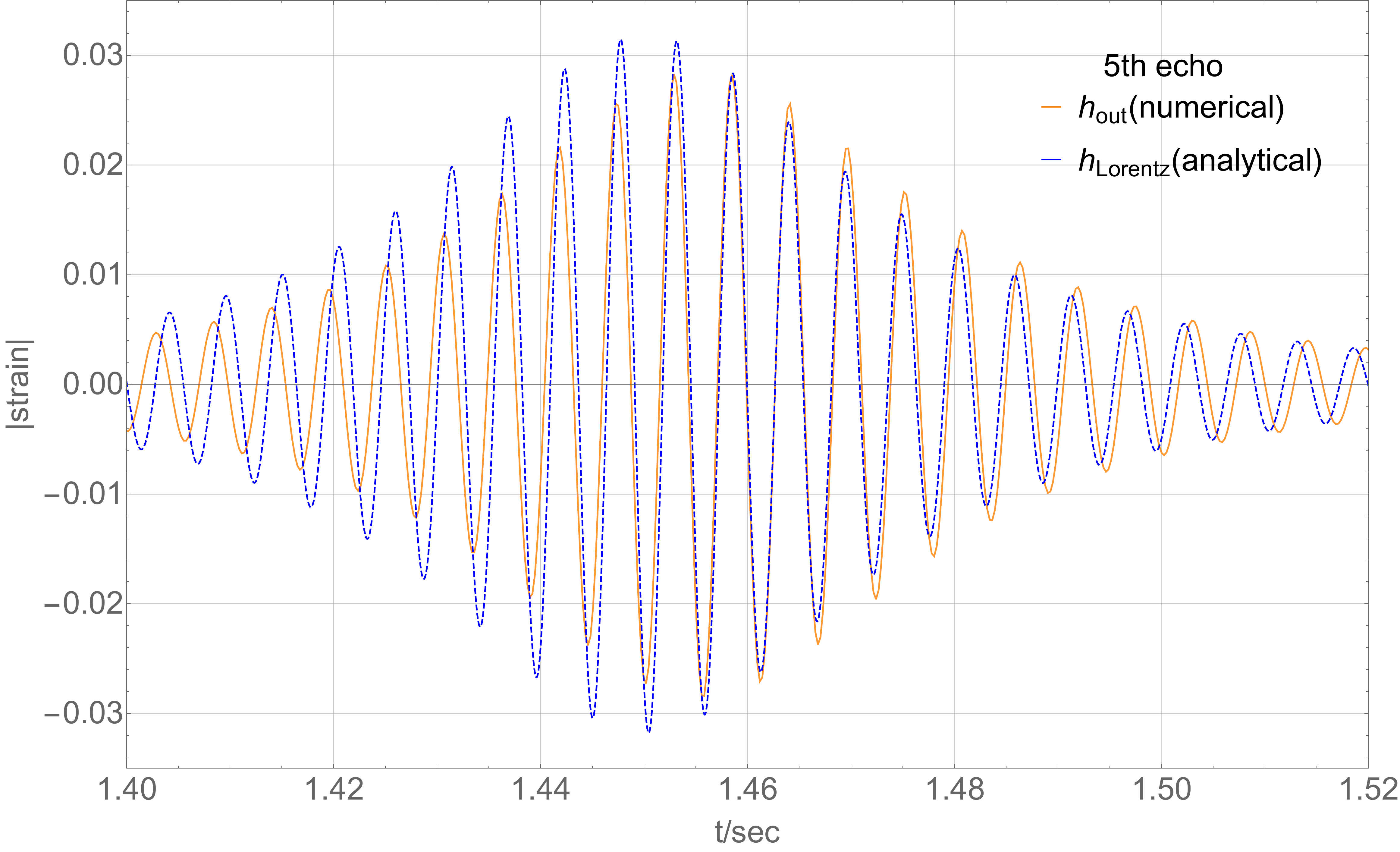}
\includegraphics[width=0.3 \textwidth]{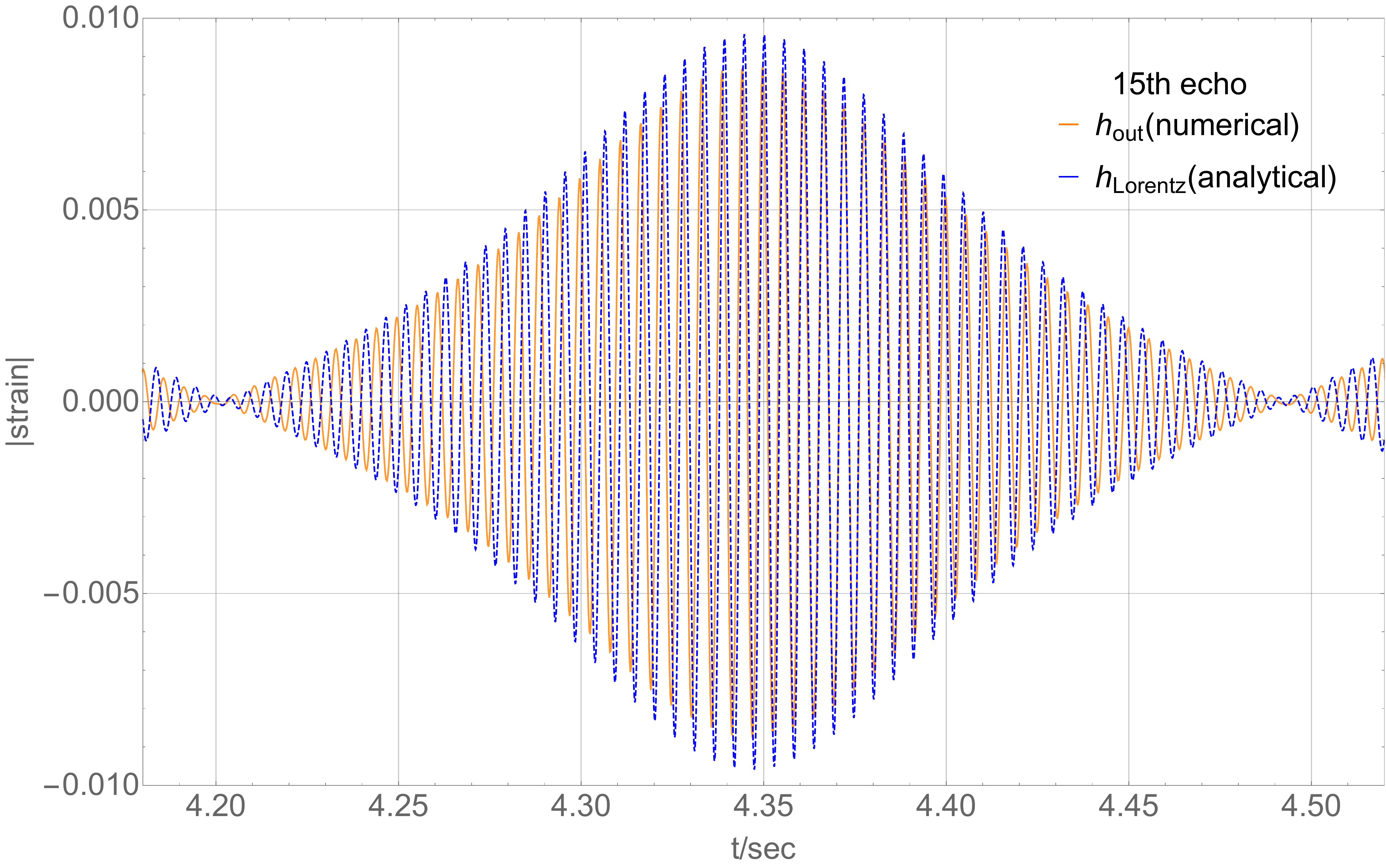}
\caption{\label{echo} The single real time echoes from the numerical geometric optics approximation (applied to LIGO template), as well as the analytic Lorentzian model. The amplitudes match well, while phases are hard to predict due to rapid oscillation over long time. we shift the numerical solution for each echo to match the phases around the peak.}
\end{figure*}

It is easy to understand this behavior analytically. Since only modes with $|\tilde{\omega}| \lesssim T_{\rm H}$ survive for many echoes, the denominator of Eq. (\ref{qnm_lorentz}) is approximately constant, and can be factored out of the sum. The rest of sum can be decomposed into two geometric series, and has a closed form:
\begin{equation}
    h_{\rm out}(t) \simeq \frac{i e^{- {ia t \over M r_+}}\cos\left(\frac{\pi t}{2x_0}\right)\sinh\left(\frac{\pi t}{8x_0^2 T_{\rm H}}\right)}{2 x_0 \left(\frac{a}{M r_+} - \omega_*\right)\left[\cos\left(\frac{\pi t}{x_0}\right)- \cosh\left(\frac{\pi t}{4x_0^2 T_{\rm H}}\right)\right]},
\end{equation}

At the peak of the $k$-th echo $t= k \times \Delta t_{\rm echo}$, corresponding to the  echo, the cosines becomes $\pm 1$ and thus the echo amplitudes can be further simplified:
\begin{eqnarray}
|h_{\rm out}(t=k \times  \Delta t_{\rm echo})| \propto \frac{1}{ \sinh\left(\frac{\pi t}{8x_0^2 T_{\rm H}}\right)},
\end{eqnarray}
which indeed, as we see in Fig. \ref{compare}, transitions from $1/t$ to exponential decay after:
\begin{equation}
k_{\rm tran} \sim \frac{8 x_0^2 T_{\rm H}}{\pi \times 2|x_0|} = -\frac{2\ln(\gamma |\tilde{\omega}|)}{\pi^2} \simeq 19,
\end{equation}
echoes, for $\gamma \tilde{\omega} = {\cal O}(10^2)$ rad/s. 
Heuristically, we can see that summing over many QNMs is responsible for the early power-law decay. However, since higher QNMs leak faster, the late-time behavior for $k \gtrsim 20$ is dominated by the least-damped QNM, which would decay exponentially. 

We can also look at the behavior around the peak of each echo, for $ k \ll k_{\rm tran}$:
\begin{equation}
h_{\rm out}(t) \propto \sum_k \frac{(-1)^k k e^{- {ia t \over M r_+}}}{\left(t- k \Delta t_{\rm echo}\right)^2+ \left(\Delta t_{\rm echo}\over \pi\right)^2 \left( k \over k_{\rm tran}\right)^2}.     
\end{equation}
In other words, the amplitude of the first $\sim 20$ echoes can be well-approximated by a Lorentzian function {\it in time}, where the ratio of echo width $\sigma_{\rm echo}$ to echo spacing $\Delta t_{\rm echo}$ is given by:
\begin{equation}
    \frac{\sigma_{\rm echo}}{\Delta t_{\rm echo}} = \frac{1}{\pi}\left( k \over k_{\rm tran}\right).
\end{equation} 
We see that the echoes are sharper initially, but start to merge for $k \sim k_{\rm tan}$, which is where we effectively transition to a single damped QNM. 

This behavior can be seen in Fig. \ref{echo}, where we plot the  1st, 5th and 15th echo, individually. Again, we see that the amplitude decays as $1/k$ for for the k-th echo, while its width grows as $k$. Here, we slightly shift the numerical solution to match the phases around the peaks, since it is hard to predict phases correctly due to rapid oscillations.

\begin{figure}
\includegraphics[width=0.4\textwidth]{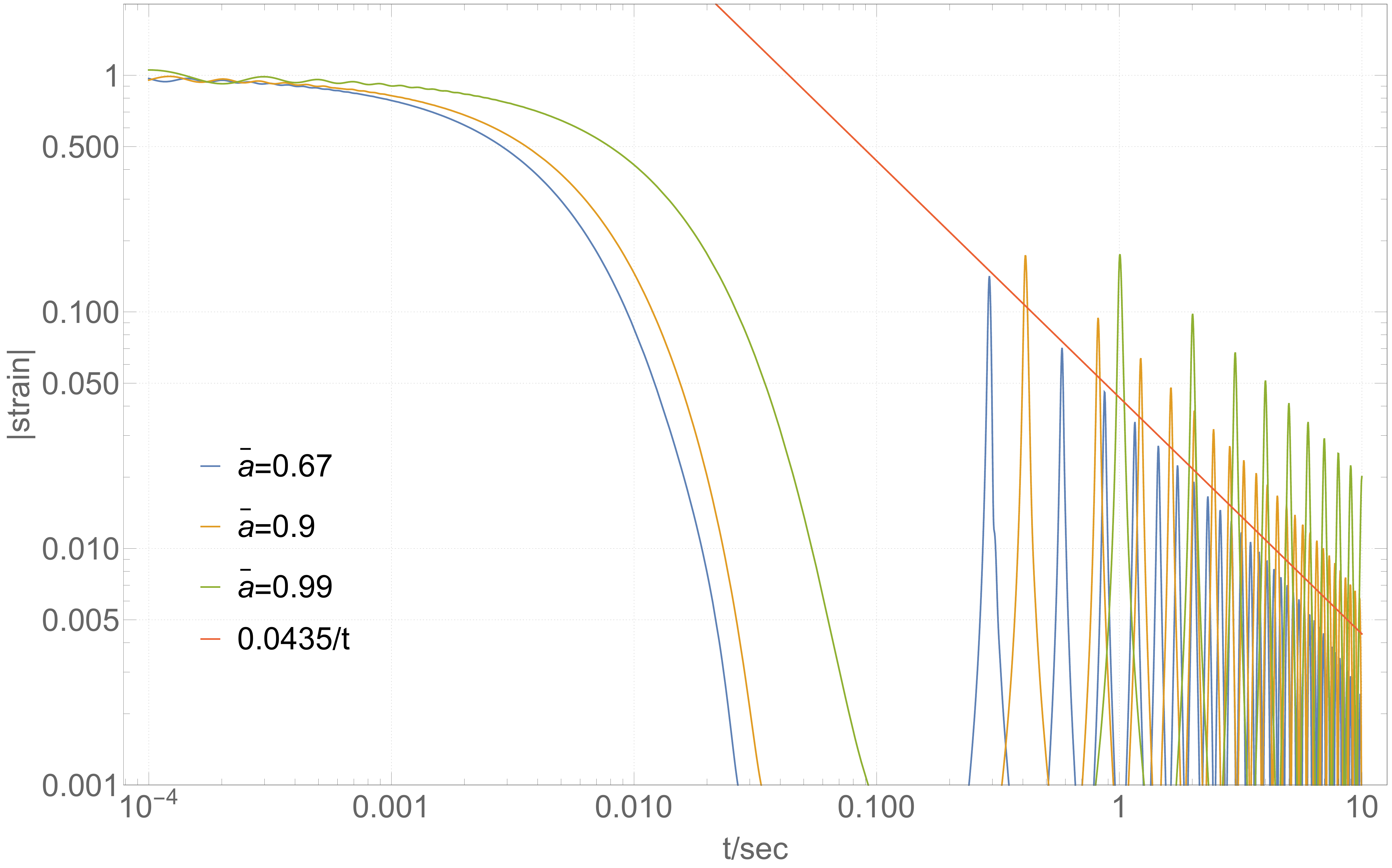}
\caption{\label{spin} The real time echoes from the Lorentzian model for  different spins with the same mass as GW150914, and $\gamma \sim 1$. Similar to Fig. \ref{compare}, we see that they all decay as power laws at the early times.}
\end{figure}

Given the accuracy of the Lorentzian model in capturing echo properties, we can apply it to different spins, starting with their fundamental ($n=1$) classical QNM for $s=-2, l=2$ and $m=2$ from the public source \footnote{\href{https://centra.tecnico.ulisboa.pt/network/grit/files/ringdown/}{https://centra.tecnico.ulisboa.pt/network/grit/files/ringdown/}}, fixing the mass to $67~ M_\odot$. We see in Fig. \ref{spin} that quantum BHs with higher spins have longer $\Delta t_{\rm echo}$, and slightly higher amplitudes (normalized to their classical QNM amplitude), while they all show a similar power law decay at the early times. 

\begin{figure}
\includegraphics[width=0.4\textwidth]{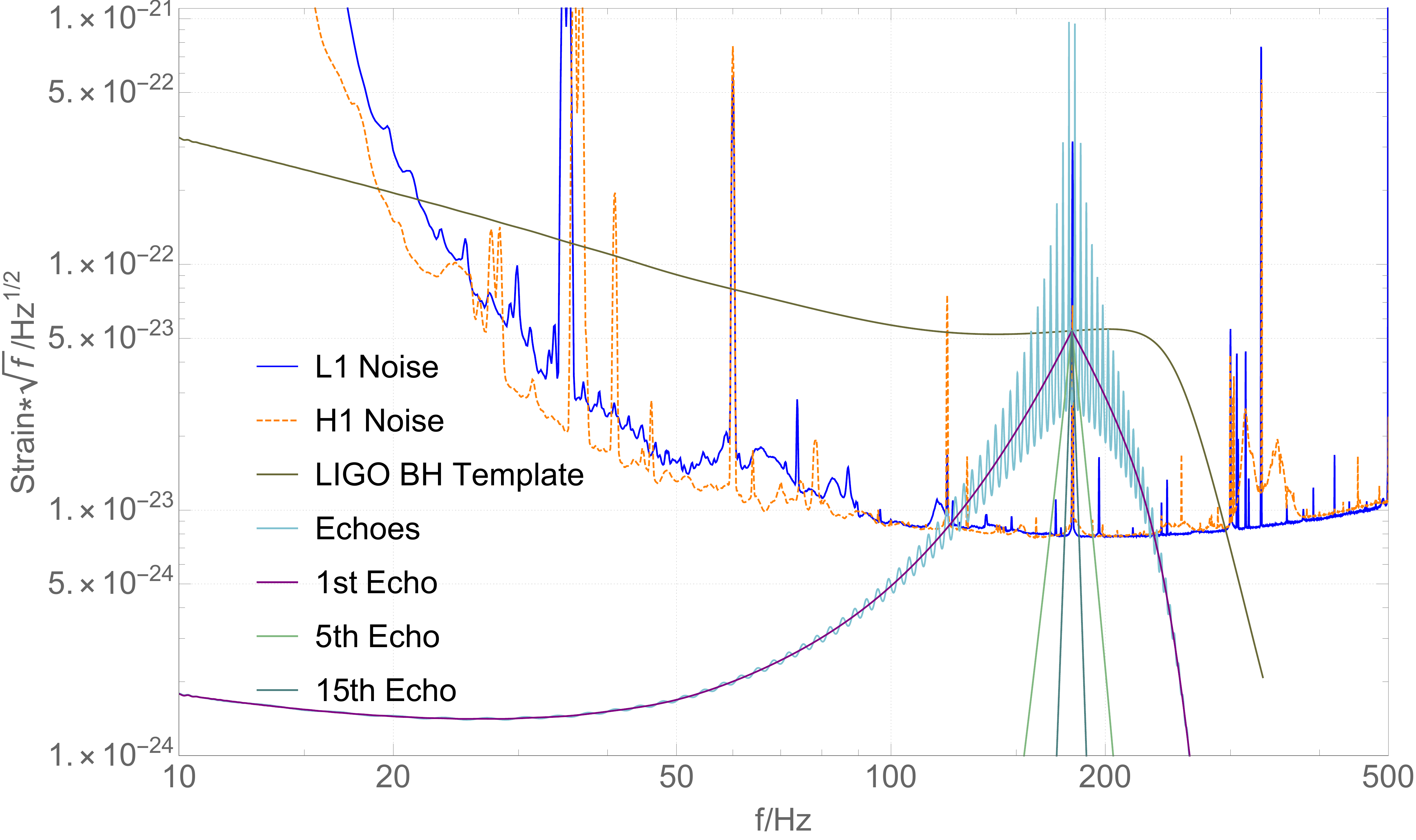}
\caption{\label{frequency} The echoes in the frequency domain compared to LIGO Hanford and Livingston noise around GW150914. Amplitude for the main, as well as expectation for the first, fifth, fifteenth, and all the echoes are shown. All the echo signals center at $\tilde{\omega} \simeq 0$ or $\omega \simeq \frac{a}{M r_+}$ as expected. Note that echo amplitudes would be lower by a factor of $\sim 3$, if we instead use the Lorentzian model in Fig. \ref{compare}, and (approximately) fix the main event amplitude. }
\end{figure}

\subsection{On detectability of Boltzmann echoes}

To get a sense of the detectability of the our Boltzmann echoes from a quantum BH, we study the signal to noise ratios (SNRs) of the echoes that we obtain from GR template of GW150914, and compare it to that of the binary black hole merger event. Here, $\rm SNR^2 \equiv  \sum_{f} \frac{ | \hat{h}_{f}| ^2}{{\sigma_{f}}^2}$, where $\hat{h}_{f}$ is strain in the frequency domain, and $\sigma_{f}$ is the detection noise of LIGO. Fig. \ref{frequency} shows the strains in the frequency domain. Comparing the LIGO noise \footnote{\href{https://www.gw-openscience.org/events/GW150914/}{https://www.gw-openscience.org/events/GW150914/}} curves with the echoes illustrates that they stand out of the noise around 100Hz to 300Hz, and reach the biggest amplitude at $\tilde{\omega} \sim 0$, where Boltzmann reflectivity reaches a maximum.

\begin{figure}
\hspace*{-.5cm}
\includegraphics[width=0.5\textwidth]{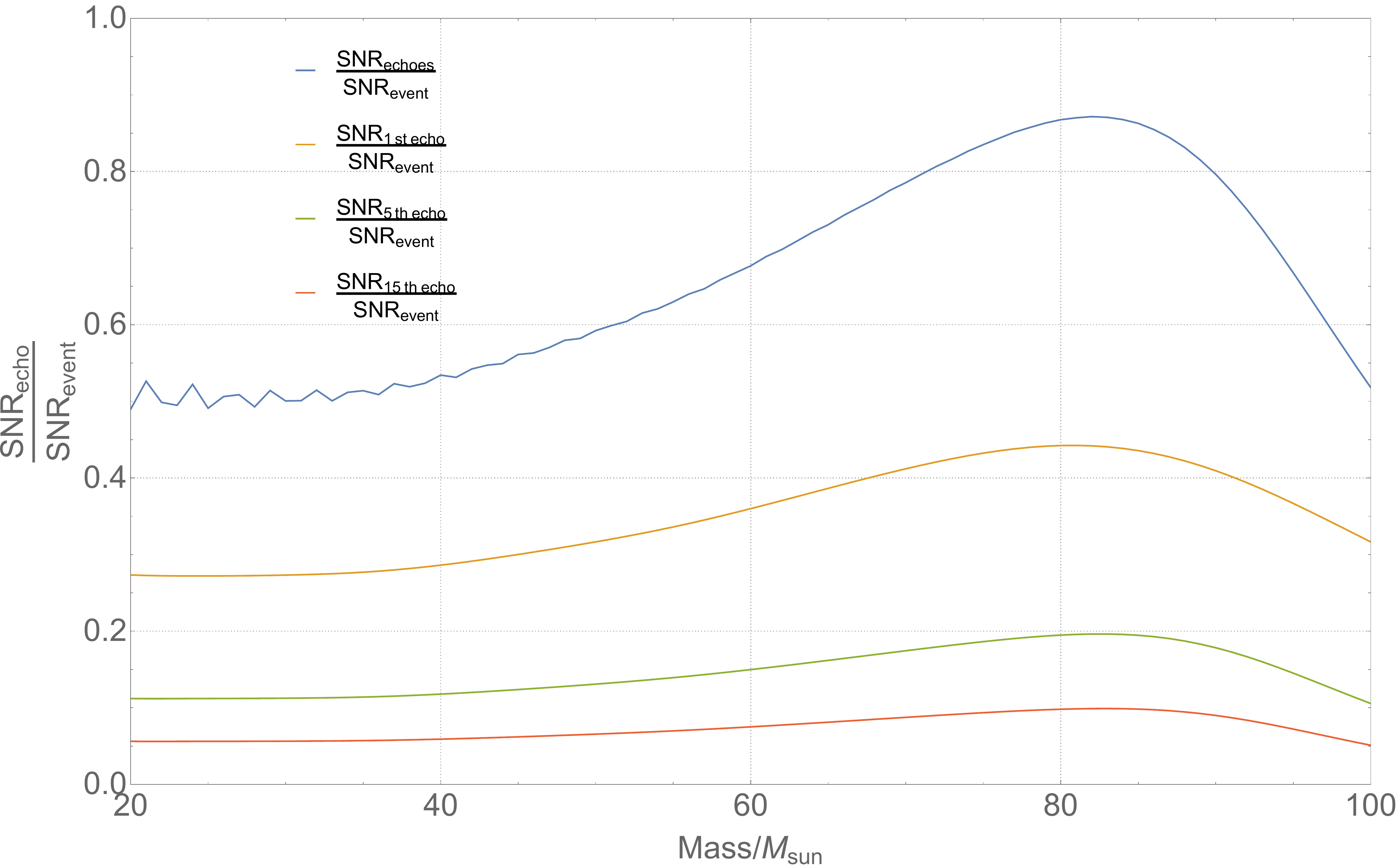}
\caption{\label{snr} SNR of echoes over main event with the shifted mass (using GW150914 but shifting the data to effectively change the mass). Note that this ratio would be lower by a factor of $\sim 3$, if we instead used the Lorentzian model in Fig. \ref{compare}.}
\end{figure}

Using LIGO noise (combining Hanford and Livingston detectors), we calculate the expected ratio ${\rm SNR}_{\rm echoes}/{\rm SNR}_{\rm event}$, which is shown in Fig. \ref{snr}. Here we red(blue)shift the LIGO BH template as in Fig. \ref{frequency} to compare SNRs with different effective masses. The ratio peaks around 85 $M_\odot$, close to the GW150914 event final mass of 67$~M_\odot$.

Some words of caution are in order: First, we should remind the reader that all the calculations presented here use linear perturbation theory, while the initial conditions of binary black hole mergers are clearly non-linear. This uncertainty in initial conditions can be see by the difference in the amplitude at $t \to 0$ between the Lorentzian and numerical model in Fig. \ref{compare}: The ratio of 1st echo to main event peak is $0.44$ for the numerical model, while it is $0.13$ for the Lorentzian model. This factor of $\sim 3$ difference reflects the uncertainty that arises from (lack of) proper nonlinear modeling of the initial conditions. 

Another point is that, any echo model would have additional free parameters, such as $\gamma$ or echo phases, which need to be fitted for, and effectively reduce the significance of echoes, if one properly accounts for the look-elsewhere effects.

\begin{figure}
\includegraphics[width=0.4\textwidth]{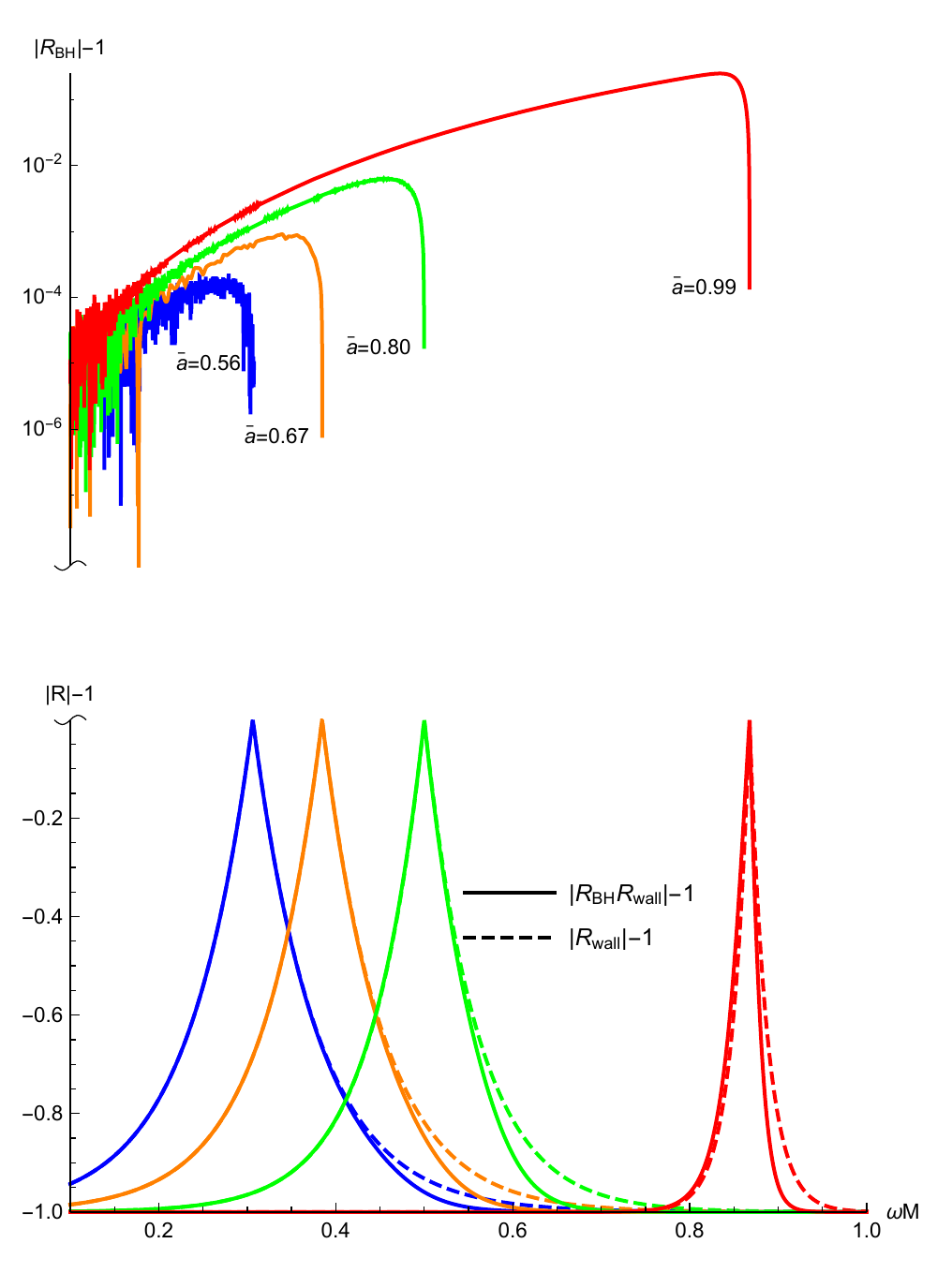}
\caption{\label{superradiance} Comparison between reflectivity of angular momentum barrier for Kerr BHs (top panel) and Boltzmann reflectivity (bottom panel, dashed curves), and their product (bottom panel, solid curve) for different spins. We see that superradiance is highly suppressed by the Boltzmann factor.}
\end{figure}

\section{Ergoregion Instability?}

\label{sec4}
Potential ergoregion instability has been a concern for the models of exotic compact objects (ECOs), since a perfectly reflective wall with the angular barrier potential catching the modes in the superradiance frequency range might lead to instability for all the spinning ECOs \cite{Maggio:2017ivp,Maggio:2018ivz}, in contradiction with observations \cite{Narayan:2013gca,Barausse:2018vdb}. However, the quantum BHs that follow the fluctuation-dissipation theorem do not suffer from this instability since the superradiance is highly suppressed because of the Boltzmann reflectivity. This is illustrated in Fig. \ref{superradiance}: the top panel is the standard superradiance for BHs and the bottom plots reflectivity of quantum BHs (for one reflection), which never exceeds 1 for different spins.

\section{Conclusion}

\label{sec5}
In a companion paper \cite{Oshita:2019sat}, we advanced independent arguments for why classical horizons must be replaced with stretched horizons with Boltzmann reflectivity for quantum BHs, which are only perfectly absorbent for frequencies much bigger than that of Hawking photon. Using the concrete boundary conditions that result from the fluctuation-dissipation theorem in \cite{Oshita:2019sat},  we analyzed the QNMs of quantum BHs analytically, and confirmed that the resulting predictions are consistent with numerical real-time echoes (in linear perturbation theory), that result from mergers of binary BHs. The echo waveforms are computed, both from geometric optics approximation and a sum over QNMs (which has a closed analytic form). 

Considering the uncertainty in modeling the nonlinear initial conditions of Boltzmann echoes and LIGO noise properties, we predict that the SNR for first (all) echo(es) is 13-44\% (24-82\%) of the SNR for the main binary merger event.

Finally, we argue that with the efficient absorption from the Boltzmann factor, ergoregion instability is suppressed for all spins.

\begin{acknowledgements}
We thank Vitor Cardoso, Elisa Maggio, Rafael Sorkin, Huan Yang, Aaron Zimmerman, Bob Holdom and Ren Jing for helpful comments and discussions. We also thank all the participants in our weekly group meetings for their patience during our discussions. This work was supported by the University of Waterloo, Natural Sciences and Engineering Research Council of Canada (NSERC), and the Perimeter Institute for Theoretical Physics. N. O. is supported by the JSPS Overseas Research Fellowships. Research at the Perimeter Institute is supported by the Government of Canada through Industry Canada, and by the Province of Ontario through the Ministry of Research and Innovation.
\end{acknowledgements}

\bibliography{main}
\appendix

\end{document}